\documentclass[review,authoryear,sort&compress]{elsarticle}

\usepackage[english]{babel}
\usepackage[utf8]{inputenc}%
\usepackage[T1]{fontenc}%

\usepackage{geometry}%

\usepackage{dsfont}

\usepackage{amssymb}
\usepackage{amsthm}

\usepackage{amsmath} 

\usepackage{commath} 

\usepackage{rotating}
 
\usepackage{enumerate} %

\usepackage{mathtools}%
  
\usepackage{siunitx} 

\usepackage{gensymb}    
 
\usepackage{booktabs} 

\usepackage{pgfplots}  
\usepackage{pgfplotstable}  
\pgfplotsset{compat=newest}
\usepackage{xcolor}
\usepackage{tikz} 
\usetikzlibrary{plotmarks,backgrounds,patterns,angles,quotes}
\usetikzlibrary{external}
\usepgfplotslibrary{groupplots} 
\usetikzlibrary{spy}
\tikzset{external/only named=true}

\usepackage{hyperref}

\usepackage[noabbrev,capitalize]{cleveref}
\crefname{equation}{}{}
\Crefname{equation}{}{}

\theoremstyle{plain}

\theoremstyle{plain}

\theoremstyle{plain}

\theoremstyle{plain}

\theoremstyle{plain}

\DeclareMathOperator{\dx}{\,dx}
\DeclareMathOperator{\dt}{\,dt}
\DeclareMathOperator{\ds}{\,ds}

\makeatletter
\def\ps@pprintTitle{%
   \let\@oddhead\@empty
   \let\@evenhead\@empty
   \let\@oddfoot\@empty
   \let\@evenfoot\@oddfoot
}
\makeatother

\begin{document}

\begin{frontmatter}

\title{Optimal Control of Droplets on a Solid Surface using Distributed Contact Angles}

\author[TUB]{Henning Bonart\corref{cor}}
\ead{henning.bonart@tu-berlin.de} 
\cortext[cor]{Corresponding author}

\author[UKO]{Christian Kahle}

\author[TUB]{Jens-Uwe Repke}

\address[TUB]{Technische Universit\"at Berlin, Process Dynamics and Operations Group,\\Straße des 17. Juni 135, 10623 Berlin, Germany}

\address[UKO]{Universit\"at Koblenz-Landau, Universit{\"a}tsstra{\ss}e~1, 56070 Koblenz, Germany}

\begin{abstract}
Controlling the shape and position of moving and pinned droplets on a solid surface is an important feature often found in microfluidic applications.
However, automating them, e.g., for high-throughput applications, does rarely involve model-based optimal control strategies. 
In this work, we demonstrate the optimal control of both the shape and position of a droplet sliding on an inclined surface. 
This basic test case is a fundamental building block in plenty of microfluidic designs. 
The static contact angle between the solid surface, the surrounding gas, and the liquid droplet serves as the control variable.
By using several control patches, e.g., like done in electrowetting, the contact angles are allowed to vary in space and time.
In computer experiments, we are able to calculate mathematically optimal contact angle distributions using gradient-based optimization.
The dynamics of the droplet are described by the Cahn--Hilliard--Navier--Stokes equations.
We anticipate our demonstration to be the starting point for more sophisticated optimal design and control concepts.
\end{abstract}

\begin{keyword}
Droplets, optimal control, contact angle, phase field model, electrowetting, lab-on-a-chip
\end{keyword}
 
\end{frontmatter}

\section{Introduction}\label{sec:intro}
In many droplet-based microfluidic processes and applications the precise shape and position of the droplets over time play a significant role for the performance of the device.
One example is the transport of droplets in a lab-on-a-chip device.
Here, the droplets get merged, split and mixed at specific positions on the chip~\cite{Pollack2002}.
Thereby, the shape of the droplets influences the heat and mass that is exchanged with the solid surface and the surrounding fluid phase~\cite{Al-Sharafi2018}.
A second example are optical applications where liquid droplets act as flexible lenses with continuous refraction index ranges.
By adjusting their shape the curvature and hence the focal length can be precisely tuned~\cite{Hou2007}.
In all these examples, the shape and position of the droplets (or gas-liquid interface) is regulated by adjusting the contact angles of the solid surface dynamically.
However, these microfluidic devices rarely incorporate model-based optimal control strategies.

Despite the impact of optimal control in industries like aviation, automotive, and chemical, only very few articles describe optimal control within the scope of droplet-based microfluidics.
In~\cite{Laurain2015} the control of the footprint and shape of a static droplet was presented.
The position of a moving droplet and its shape was considered in~\cite{Antil2017} in absence of gravity.
\cite{Fumagalli2017} presented work on the position of the gas-liquid interface of rising liquid in a capillary.
However, the simultaneous control of both the shape and position of a droplet to fulfill a target in a mathematically optimal way is absent.

Therefore, we present the surface based control of both the shape and position of droplets.
The static contact angle between the solid surface, the surrounding gas and the liquid droplet serves as the control variable.
The dynamics of the droplet and the gas-liquid interface are described by the Cahn--Hilliard--Navier--Stokes equations.
We refer to \cite{BonKR-OPTMCL-PAMM2019,BonKR-MCLOPT-ana} for a detailed description of the mathematical setup. 
The optimal control problem is solved using a quasi-Newton method. Gradients are derived using adjoint calculus.

The capabilities of our approach are demonstrated using two test cases:
The first one is a simple but intriguing demonstration to introduce the concept.
Here we consider  a wetting and receding droplet on a horizontal surface.
The second, more complex example is inspired by the work reported in~\cite{Al-Sharafi2018}, where the heat transfer into a sliding and pinned droplet is characterised, and~\cite{tMannetje2014}, where drops are trapped due to steep changes in the contact angle.
We assume, that our approach and demonstration initiates further research on optimal control of droplet-based microfluidics.

The remainder of the paper is structured as follows.
We discuss the control problems and our proof-of-concept in~\cref{sec:problems}.
It follows a description of the control and the optimization problem in~\cref{sec:optimization} and the forward model in~\cref{sec:model}. 
The results of the multiple control problems including the sliding droplet are given in~\cref{sec:results}.
A summary and conclusion is the subject of the last section.

\section{Description of Test Cases}\label{sec:problems}
To show the applicability of our approach, we consider several optimal control problems.
Thereby, we always aim to optimize both the shape and position of the droplet over a given time horizon.
In \Cref{ssec:p:horizontal} and \Cref{ssec:p:inclined} we describe two experimental setups under consideration and
in \Cref{ssec:ConsideredControls} we state four approaches the achieve optimal control in these settings that will be tested in this work.

\subsection{Droplet on a Horizontal Surface} \label{ssec:p:horizontal}
\begin{figure}[!h]
	\centering
	\begin{tikzpicture}[scale=3.0]
	\filldraw[thick, rotate=0, fill=black!10] (0.25, 0) node(r1){} arc(180:0:0.25)node(r2){};
	\draw[thick, rotate=0, dashed] (0.36, 0.0) node(r1){} arc(225:-45:0.20)node(r2){};
	\draw[thick] (0.0,0) node (a){} -- (1.0,0) node (b){};
	\fill[thick,pattern=north west lines] (0.0,0) node (a){} -- (1.0,0) node (b){} -- (1.0,-0.05) node (c){} -- (0.0,-0.05) node (d){} -- (0.0, 0);
	\draw[<-] (0.22, 0.10) -- (0.0, 0.25) node[above, font=\fontsize{8pt}{8pt}\selectfont, align=center]{initial shape/position\\$\varphi_0$};
	\draw[<-] (0.7, 0.25) -- (0.9, 0.35) node[above, font=\fontsize{8pt}{8pt}\selectfont, align=center]{desired shape/position\\$\varphi_d$ at  time $t$};
	\draw[thick] (0, -0.1) -- ++(0.33, 0) node (r){};
	\draw[thick] (r) -- ++(0.33, 0) node (r){};
	\draw[thick] (r) -- ++(0.33, 0) node (r){};
	\draw[<-] (1, -0.1) -- (1.2,0) node[above, font=\fontsize{8pt}{8pt}\selectfont, align=center]{control patches/\\electrodes};
		\draw [thick,decorate,decoration={brace,amplitude=4pt,mirror,raise=6pt}] (a)++(0,-0.08) -- ++(1,0) node [black,midway, below, yshift=-10pt, font=\fontsize{8pt}{8pt}\selectfont, align=center] {Control with $\theta(x, t)$};

\end{tikzpicture}
	\caption{Illustration of the optimal control problem of the symmetric, cap-shaped droplet on a horizontal surface.}
	\label{fig:res:cap_setup}
\end{figure}
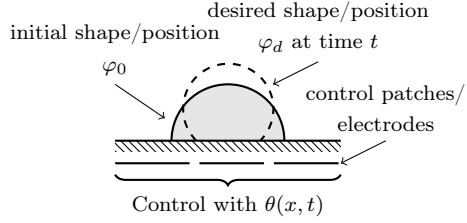

We start by revising one of the most simplest but still highly intriguing demonstrations of electrowetting.
The simplified setup is shown in~\cref{fig:res:cap_setup}.
A single droplet is placed on a horizontal surface with an equilibrium contact angle of $\theta_{eq} = \SI{90}{\degree}$.
The solid and dashed line represent the initial and desired shape and position of the droplet.
The control patches, or electrodes, below the surface are used to dynamically change the actual contact angle $\theta$.
In this way, the spreading length of the cap-shaped droplet can be precisely adjusted.
In an optical application for example this could  lead to a modulation of the refractive index.
A further example would be the adaption of the heat transfer from the solid surface into the droplet.
As we are not concerned here with the technical implementation of the contact angle modification we refer to the recent text book by~\citet{Mugele2018} for an extensive description.

Adjusting the voltage in the patches until the droplet reaches a specific shape and position just by trial-and-error and experience can be cumbersome.
This is especially true in complex applications where inertial forces of the droplet or the surrounding phase might lead to ripples or instabilities.
Therefore we do not directly specify a contact angle $\theta$ to enforce a specific shape of the droplet.
In fact, we directly incorporate the desired shape and position of the droplet into an optimal control problem.
The details of this problem  will be further described in~\cref{sec:optimization}.

\subsection{Pinning a Sliding Droplet}\label{ssec:p:inclined}
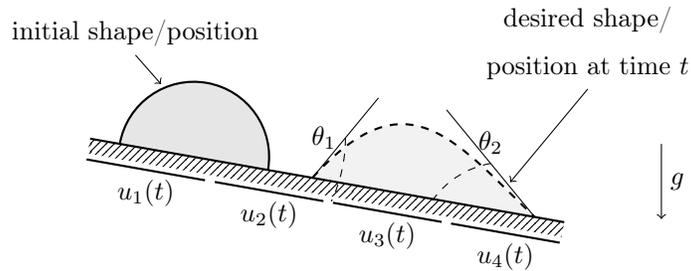
\begin{figure}[h]
	\centering
	\tikzsetnextfilename{Setup}
\begin{tikzpicture}[scale=4.0]
	\filldraw[thick, rotate=-10, fill=black!10] (0.0, 0) node(r1){} arc(180:0:0.25)node(r2){};
	\draw[<-, rotate=-10] (0.10, 0.22) -- (0.0, 0.3) node[above, align=center]{initial shape/position};
	\filldraw[rotate=-10,thick, dashed, fill=black!05] (0.65,0) node[](r1){} .. controls (0.9,0.4) and (1.1,0.2) .. (1.4,0) node(r2){};
	\draw[<-, rotate=-10] (1.3, 0.10) -- (1.5, 0.45) node[above, align=center]{desired shape/\\position at time $t$};
	\draw[rotate=-10, thick] (-0.1,0) -- (1.5,0) node(b){};
	\fill[rotate=-10,thick,pattern=north east lines] (-0.1,0) -- (1.5,0) -- ++(0,-0.05) -- (-0.1,-0.05) -- ++(0, 0.05);
	\draw[rotate=-10,thick] (-0.1, -0.07) -- node[midway,below] {$u_1(t)$}++(0.4, 0) node (r){};
	\draw[rotate=-10,thick] (r) -- node[midway,below] {$u_2(t)$}++(0.4, 0) node (r){};
	\draw[rotate=-10,thick] (r) -- node[midway,below] {$u_3(t)$}++(0.4, 0) node (r){};
	\draw[rotate=-10,thick] (r) -- node[midway,below] {$u_4(t)$}++(0.4, 0) node (r){};
	\draw[->] (1.8,0)-- node[right]{$g$}++(0.0,-0.25);
	\draw[rotate=-10] (0.65,0) --node[midway](c1){} ++(0.17,0.3);
	\draw[dashed] (c1) arc(0:-25:0.5);
	\node[left] at (c1) {$\theta_1$};
	\draw[rotate=-10] (1.4,0) --node[midway](c2){} ++(-0.35,0.3);
	\draw[dashed] (c2) arc(100:145:0.3);
	\node[above] at (c2) {$\theta_2$};
\end{tikzpicture}
	\caption{Physical setting of the optimal control problem.}
	\label{fig:setup_intro} 
\end{figure}

The example above can be extended to numerous more complex applications.
Exemplarily, we discuss a liquid droplet placed on an inclined solid surface.
This example is inspired by the work reported in~\cite{Al-Sharafi2018}, where the heat transfer into a sliding and pinned droplet is characterised, and~\cite{tMannetje2014}, where drops are trapped due to steep changes in the contact angle.
Here, we want to accomplish a similar pinning of the droplet, i.e., we want to specify a desired position of the droplet.
In addition, we want the droplet to have a specific shape at a specific instance in time.
In~\cref{fig:setup_intro}, the general physical setting of the problem is illustrated.
A single droplet (solid line) is placed on an inclined surface with an equilibrium contact angle of $\theta_{eq} = \SI{90}{\degree}$.
The dashed line represents the desired shape and position at some instance in time.
It is clear, that if no control action is taken, the droplet will slide down the surface driven by gravity.
However, with the help of the patches $u_1$ to $u_4$ we want to control the advancing and receding contact angles $\theta_1$ and $\theta_2$ while the droplet slides along the surface.
In this way, we would be able to impose the desired shape and position at a specific time $t$ (light gray).

Apparently, it would be highly cumbersome to find control actions for the patches $u_1$ to $u_4$ by extensive trial-and-error.
Furthermore, no guarantee can be given, that the control action based on experience would be in some sense optimal.
Again we formulate an optimal control problem and incorporate the desired shape into the optimization functional.

\subsection{Control Actions}
\label{ssec:ConsideredControls}
Four different types of control actions will be discussed in this work.
We start with the most simplest type and successively approach the proof-of-principle of an optimal control of both the shape and position of a droplet.

\begin{description}
\item[A: Constant in Space and Time] 
In this setting, the control is given by one contact angle, that is constant with respect to space and time.
\item[B: Spatially Adapted] Here, we separate the surface into equally distributed stripes (or electrodes) and solve the optimal control problem.
Each strip can feature a different but constant contact angle.
Again, as the contact angles are constant in time (but not in space), we aim to calculate the optimal design of a surface.
\item[C: Temporal Switchable] For the next step we consider a real active control problem.
That means, that this result can not be implemented in applications without some kind of active control mechanism like electrowetting.
In this example, we allow the spatially homogeneous contact angle to switch at a few discrete instances in time.
This correspondences to electrowetting with a single electrode.
\item[D: Changeable in Space and Time] 
Finally, we consider the most general optimal control problem. This in fact is
needed to consider the sliding droplet case explained in \Cref{ssec:p:inclined}.
Therefore, we combine the second and third example.
The surface is separated into equally distributed stripes. 
Each stripe is allowed to switch its contact angle at a few discrete instances in time.
\end{description}
In the following we consider the control actions A and B as passive controls or manufactured controls, because the control action is 
predefined offline by the manufacturing of the solid surface, while we call the control types C and D as active controls,
as they are applied online, while the actual dynamics happens.

\section{Construction of the Design/Control}\label{sec:optimization} 
In this section, we describe the optimization problem that we investigate to obtain optimal controls, i.e. optimal distributions of contact angles with respect to time and space. 
The following modeling is independent of the actual description of the droplet. 
We only assume, that there is some measure to quantify the quality of a droplet with respect to some objective. 
We give examples at the end of this section.
We also refer to our complementary works~\cite{BonKR-OPTMCL-PAMM2019,BonKR-MCLOPT-ana} for an extensive description of the mathematical details. 

We assume, that we can influence the contact angle at the contact line with a suitable control mechanism (i.e., electrowetting).
To state the optimal control problem that defines our controller, we assume, that we have some general description $\varphi$ of the droplet.
Moreover, we have some abstract control variable $u$ together with an embedding operator $B$ such that $Bu$ models the actual applied distribution of contact angles in space and time. 
We finally assume, that for any control $u$ we can calculate a droplet $\varphi(u)$ utilizing some kind of forward model.
The considered optimal control problem is given as
\begin{equation}
\tag{P}
\label{prob:O:P}
\begin{aligned}
	\min_{\varphi,u} J(\varphi,u)& = 
	\frac{1}{2}\int_{0}^T\int_\Omega \omega(t) |\varphi(t,x)-\varphi_d(t,x)|^2\dx\dt
	+ \frac{\alpha}{2}\int_0^T\int_{\partial\Omega} |(Bu)(t,x)|^2 \dx\dt \\
	\mbox{subject to } & \varphi  = \varphi(u),\\
	& \cos \theta_{\max} \leq Bu  + \cos (\theta_{eq})   \leq \cos \theta_{\min},	
\end{aligned}
\end{equation}
where $\varphi_d$ represent the desired shape and position of the droplet, $\alpha>0$ denotes a chosen regularization parameter and $\theta_{eq}$ is the equilibrium contact angle of the solid surface without any control action.
The actual objective that we encode in the first addend of $J$ is the tracking of a given evolution $\varphi_d$ over the time horizon $[0,T]$ in the least-squares sense by minimizing their mismatch.
We additionally add a time depending weight $\omega(t) \geq 0$ that can be used to pronounce the mismatch in some parts of the time horizon.
The second addend penalises strong control actions and can be used to minimize the controller energy.
The angles $0 < \theta_{\min} < \theta_{\max} < \pi$ are given minimum and maximum contact angles. Note that cosine is
monotonically decreasing in this range, which leads to the unexpected direction of the inequalities.
This can encode physical restrictions, i.e., the contact angle can not be negative, or technical issues, i.e., the control system or technique can not reach contact angles above a certain value.
We have chosen this constraint to closely resemble the Young-Lippmann equation for basic electrowetting~\cite{Mugele2018}.
For example, if electrowetting is applied to manipulate the contact angle, the maximum accessible contact angle is equal to the equilibrium contact angle.
Note, that the precise form of this objective $J$ is not important for the conceptual approach and one might use other models to quantify the quality of a given droplet or include further restrictions.
For a general introduction into the field of optimization with constraints given by partial differential equations we refer to \cite{TroelOptiSteu,Hinze2009}.

\subsection{Model for Control Action $u$}
\label{ssec:control}
We model the control $Bu$ by a linear combination of fixed and given control actions. 
Here the control variable $u = (u_{rs})_{r = 1,\ldots,R}^{s=1,\ldots,S}$ represents the amplitudes of $R$ given control actions $(g_r(t))_{r=1}^R$ that purely depend on time and $S$ control actions $(f_s(x))_{s=1}^S$ that purely depend on space. 
We use the model
\begin{align*}
(Bu)(t,x) = \sum_{r=1}^R\sum_{s=1}^S u_{rs}g_r(t)f_s(x).
\end{align*}
In this way we can describe the four settings (A--D) under consideration stated in \Cref{ssec:ConsideredControls}.
To model the setting A we use $R=1$, $g_1(t) = 1$, and $S=1$, $f_1(x) = 1$, such that $Bu(t,x) = u_{11}$ describes a constant contact angle in time and space.
For the setting B, i.e. controls that depend on space, but that are constant in time, we use $R=1$, $g_1(t) = 1$ and $S\geq 1$, 
while the setting C, i.e. the same contact angle on the whole boundary, that might change over time, we use $R\geq 1$  and $S=1$, $f_1(x) = 1$.
Finally, for setting D we use $R\geq 1$ and $S\geq 1$.
Moreover, we only consider controls, that are piecewise constant with respect to time and space,  respectively.
Thus if we use $S$ controls over the time horizon $T$, we use $g_s(t) = \chi_{I_s}$.
Here, $\chi_{I_s}$ denotes the characteristic function of the interval $I_s = (\frac{s-1}{S}T,\frac{s}{S}T]$, i.e. $\chi_{I_s}(t) = 1$ if $t\in {I_s}$ and $\chi_{I_s}(t)=0$ if $t\not \in {I_s}$. 
The same structure is used with respect to space.

\subsection{Description of  the Droplet $\varphi$} 
Let us finally comment on some possible ways to describe the droplet. 
For an in-depth description of available methods we refer to~\cite{Worner2012}.
In interface tracking methods, the droplet is numerically defined
by an explicit description of the interface between
gas and liquid. Here one might construct a characteristic (or indicator) function for the droplet out of
this description and consider this function as $\varphi$ in the sense of \cref{prob:O:P}. Such a characteristic
function is directly given by volume-of-fluid methods. If a level-set method is used, we also suggest to 
create a corresponding characteristic function. In other case, the level-set-functions for $\varphi$ and $\varphi_d$
need sufficient normalization such that the mismatch $|\varphi-\varphi_d|$ is sufficiently localized in that region where 
the droplets that are described by $\varphi$ and $\varphi_d$ are different.
In phase field method, the droplet is described by a smooth indicator function similar to a characteristic function that can directly be used in \cref{prob:O:P}.
We use the latter method in this work and give a brief introduction in the next section.

\section{Forward Model}\label{sec:model} 
In this section, we introduce the model for the evolution of the droplet that is used throughout this work.
We consider a phase field model, coupling the incompressible Navier--Stokes equations with the Cahn--Hilliard equation, see \cite{AbelsGarckeGruen-CHNSmodell,2006-QianWangShen-Variational-MovingContactLine--BoundaryConditions,Jacqmin2000}. 
Compared to sharp interface methods, phase field methods replace the infinitely thin layer (interface) between gas and liquid by a transition region with  positive thickness.
Moreover, they describe the distribution of the different fluids by a smooth indicator function (phase field) and do
 not explicitly involve the location of the transition region.
It follows, that all physical properties like density or viscosity vary continuously across the interface and can be described in terms of the phase field.

Especially, the Cahn--Hilliard equation allows the contact line to move naturally on the solid surface due to a diffusive flux across the interface even if no-slip is assumed~\cite{Jacqmin2000}.
As summarized in the review by~\cite{Worner2012}, the Cahn--Hilliard--Navier--Stokes (CHNS) equations can easily handle topological changes of the interface (merging and breakup)~\cite{Anderson1998}, 
the contact line can be accurately represented~\cite{Seppecher1996} and the interface is implicitly tracked without any prior knowledge of the position.
Furthermore, one of the major advantages is, that the formulation of the surface tension force in the Navier--Stokes (NS) equation exactly conserves both the surface tension energy and kinetic energy. 
This can reduce spurious currents, which are purely artificial velocities around the interface, to the level of the truncation error even for low Capillary numbers~\cite{He2008, Jamshidi2018}.

\bigskip

The liquid droplet and the surrounding air are modeled as Newtonian, isotherm, immiscible and incompressible fluids.
In this work, the common incompressible, single-field Navier--Stokes (NS) equation is combined with the convective Cahn--Hilliard (CH) equation to describe the interface dynamics
and we use the model from \cite{AbelsGarckeGruen-CHNSmodell} for the bulk dynamics.
To describe the spatial distribution of the two phases (liquid and gas), and thus the diffuse interface between them, with a single variable, an order parameter or phase field $\varphi$ is introduced as
\begin{align*}
	\varphi(x, t) = u_l - u_g = 
		\begin{cases}
				-1\text{ for pure phase 1 or }u_l = 0\;, \\
				+1 \text{ for pure phase 2 or }u_l = 1\;, \\ 
		\end{cases}
\end{align*} 
in which $u_g$ and $u_l$ with $u_l+u_g=1$ are the volume fractions of phase 1, in the following called \textit{gas} and phase 2, in the following called \textit{liquid}.
Note that this method postulates the existence of a diffuse interface between the two phases where both phases are present. In this region both $u_l \neq 0$ and $u_g \neq 0$ holds and thus
$\varphi \in (-1,1)$.
Since $\varphi$ is assumed to be continuous across the diffuse interface, in practice we can interpret its zero-level line as a sharp separation line between gas and liquid and use this line
to present numerical results.
The further primal variables are given by the velocity field $v$, the pressure field $p$, and the chemical potential $\mu$.
In this work the thermodynamically consistent diffuse interface model for large density differences between gas and liquid proposed in~\cite{AbelsGarckeGruen-CHNSmodell} is applied
\begin{align}
\rho\partial_t v + (\rho v + J)\nabla v -\mbox{div}\left(2\eta Dv\right) + \nabla p &= -\varphi\nabla \mu + \rho g\;, \label{eq:M:1_NS1}\\
-\mbox{div}(v) &= 0\;, \label{eq:M:2_NS2}\\
\partial_t \varphi + v \nabla \varphi - b\Delta \mu &= 0\;,  \label{eq:M:3_CH1}\\
-\sigma\epsilon\Delta \varphi + \frac{\sigma}{\epsilon}W^\prime(\varphi) &= \mu\;, \label{eq:M:4_CH2}
\end{align}
which is closed with the  initial data and  boundary conditions, compare \cite{2006-QianWangShen-Variational-MovingContactLine--BoundaryConditions,GruenMetzger__CHNS_decoupled},
\begin{align}
 v(t=0) &= v_0\;,&	            v &= 0\;, \label{eqn:bcs_1}\\
\varphi(t=0) &= \varphi_0\;, &	\sigma\epsilon \nabla\varphi\cdot\nu_\Omega + \gamma_u^\prime(\varphi) &=0\;, \label{eqn:bcs_3}\\
 &&	                            \nabla \mu \cdot\nu_\Omega &= 0\;, \label{eqn:bcs_4} 
\end{align}
where $\nu_\Omega$ denotes the outer unit normal on the computational domain.
We use the abbreviations $J := -b\frac{\partial\rho}{\partial\varphi}\nabla \mu$, where $b$ denotes the constant mobility in the Cahn--Hilliard model, and $2Dv := \nabla v + (\nabla v)^t$. 
The gravitational acceleration is denoted by $g$ and $p = p^{phys} - \mu\varphi$ is a shifted pressure, where $p^{phys}$ denotes the physical pressure.
The density function is denoted by $\rho \equiv \rho(\varphi) > 0$ and satisfies $\rho(-1) = \rho_g$ and $\rho(1) = \rho_l$, with $\rho_g,\rho_l$ denoting the constant densities of the two involved fluids, i.e. gas and liquid. 
It is a linear function with respect to $\varphi$.
The viscosity function is $\eta \equiv \eta(\varphi) > 0$ and satisfies $\eta(-1) = \eta_g$ and $\eta(1) = \eta_l$, 
with $\eta_g,\eta_l$ denoting the viscosities of the involved fluids.
Here, $\rho$ and $\eta$ are given by
\begin{align*}
\rho(\varphi) := & \frac{\rho_l+\rho_g}{2} + \frac{\rho_l-\rho_g}{2}\varphi,
&
		\eta(\varphi) := & \frac{\eta_l+\eta_g}{2} + \frac{\eta_l-\eta_g}{2}\varphi.
\end{align*}
The function $W(\varphi)$ denotes a dimensionless potential of double-well type.
Here, we choose 
\begin{align}
\label{eq:M:W}
W(\varphi) := 
	\begin{cases}
		\frac{1}{4}(1-\varphi^2)^2 & \mbox{if } |\varphi|\leq 1,\\
		(|\varphi|-1)^2 & \mbox{else.}
	\end{cases} 
\end{align}
The constant $\sigma = c_W\sigma_{lg}$ denotes the surface tension between liquid and gas, where $\sigma_{lg}$ denotes the physical
value of the surface tension and the constant $c_W$ is a necessary scaling of the physical surface tension that appears from the diffuse interface approach.
For the potential \cref{eq:M:W} $c_W$  is given by  $c_W = \frac{3}{2\sqrt{2}}$, compare \cite{Bonart2019b}.
The constant $\epsilon$ is proportional to the thickness of the diffuse transition zone between gas and liquid.
For $\epsilon\to 0$, i.e. if the transition zone between gas and liquid becomes small,
at least formally a consistent sharp interface model can be recovered, see \cite{AbelsGarckeGruen-CHNSmodell,2018-XuDiHu-SharpInterfaceLimit-NavierSlipBoundary}.

Finally, the contact line energy is denoted by $\gamma$ and interpolates smoothly between the solid-liquid and solid-gas surface energies.
The expression for $\gamma$ is given by
\begin{align}
\label{eq:M:gamma}
	\gamma_u(\varphi) & = \frac{\sigma_{sl}+\sigma_{sg}}{2} + \sigma_{gl} (\cos{\theta_{eq}} + Bu )\vartheta(\varphi)
\end{align} 
where $\vartheta$ is a smooth monotonically increasing function with $\vartheta(-1) = -\frac{1}{2}$ and $\vartheta(1)=\frac{1}{2}$ such that 
$\gamma_u(-1) = \frac{1}{2}(\sigma_{sl}+\sigma_{sg} - \sigma_{gl}(\cos{\theta_{eq}} + Bu )) = \sigma_{sg}$ is satisfied by Young's law, i.e. $\sigma_{sl} - \sigma_{sg} = \sigma_{lg}(\cos\theta_{eq} + Bu)$,
and $\gamma_u(1) = \sigma_{sl}$ holds by the same arguments.
Here $\theta_{eq}$ denotes the static equilibrium contact angle between the solid and the interface and is measured in the liquid phase,
while $\sigma_{sg}$ denotes the surface tension between solid and gas, and $\sigma_{sl}$ denotes the surface tension between solid an liquid.
Further, $Bu$ denotes some change of the contact angle due to the applied control, see \Cref{ssec:control} for information on $u$ and $B$.
Note the close resemblance of our formulation of~\cref{eq:M:gamma} and the Young-Lippmann equation for basic electrowetting~\cite{Mugele2018}.
It follows, that the calculated controls $u$ can be readily implemented in electrowetting devices.

Finally we note, that an expression for $\vartheta$ can for example be derived 
based on the assumption of equipartition of energy, 
i.e., $\frac{\epsilon}{2}|\nabla \varphi|^2 \approx \frac{1}{\epsilon}W(\varphi)$~\cite{Ding2008}.
This leads to $\vartheta^\prime(\varphi) = c_W\sqrt{2W(\varphi)}$.
 
For further details about the model as well as different potentials and contact line energies we refer to~\cite{Bonart2019b}.
For applications of the model to rising bubbles see~\cite{Bonart2019b,Aland_Voigt_bubble_benchmark}, to the pinning of droplets see~\cite{Bonart2019a} and to thin liquid films see~\cite{Bonart2018a}.
Especially for a comparison with sharp interface simulations and experimental measurements without contact line motion, see \cite{TaylorFlow_SPP1506_qualitativeComparison}.

The model \eqref{eq:M:1_NS1}--\eqref{eq:M:4_CH2} can be derived purely from thermodynamic principles~\cite{AbelsGarckeGruen-CHNSmodell,2006-QianWangShen-Variational-MovingContactLine--BoundaryConditions}.
Here it is postulated, that the system in the whole domain $\Omega$ with boundary $\partial\Omega$ can be described by the following sum of kinetic energy and
Helmholtz free energy functional of Ginzburg--Landau type~\cite{Jacqmin1999}
\begin{align}
		E = \frac{1}{2}\int_\Omega \rho |v|^2\;\dx + \sigma\int_\Omega \epsilon^{-1}W(\varphi) + \epsilon|\nabla \varphi|^2\dx + \int_{\partial\Omega} \gamma\ds.
\end{align}

\subsection{The discrete scheme}
The Cahn--Hilliard--Navier--Stokes equations form a very tightly coupled and nonlinear system of four partial differential equations. 
Furthermore, the Cahn--Hilliard equation \cref{eq:M:3_CH1}--\cref{eq:M:4_CH2} involves fourth-order derivatives with respect to $\varphi$.
Compared to the Navier--Stokes equation \cref{eq:M:1_NS1}--\cref{eq:M:2_NS2}, which involve only second-order derivatives, this complicates the numerical treatment~\cite{Worner2012}. 

For a practical implementation in a finite element scheme we introduce a time grid $0 = t_0 < t_1 < \ldots<t_{m-1} < t_m< \ldots <t_M = T$ on $I = [0,T]$ with (not necessarily) equidistant step size $\tau>0$.
We further introduce a triangulation $\mathcal T_h$ of the domain into cells $T_i$, such that $\mathcal T_h = \bigcup_{i=1}^{N}T_i$ covers the domain. 
On $\mathcal T_h$ we introduce piecewise linear Lagrangian finite elements $V_1 = \mathcal{P}_1$ for $\varphi_h$, $\mu_h$ and $p_h$ and the triangular/tetrahedral Mini element $V_2 = \mathcal{P}_1\bigoplus \mathcal{B}_{1+d}$, 
denoting the space of linear polynomials enriched by a cubic/quartic bubble function, for $v_h$.
For the derivation of the weak form as well as the proof of energy stability and thermodynamic consistency we refer to~\cite{Bonart2019b}.

With respect to time, we consider a time-stepping scheme and state the system that we solve on one time instance.
Given 
$\varphi^{m-1} \in V_1$, 
$\mu^{m-1} \in V_1$, and
$v^{m-1} \in V_2$, 
find 
$\varphi^m_h \in V_1$, 
$\mu^m_h \in V_1$,
$p^m_h \in V_1$ and
$v^m_h \in V_2$,
such that for all 
$w \in V_2$,
$q \in V_1$,
$\Psi \in V_1$, and
$\Phi \in V_1$
the following equations hold
\begin{align}
   \frac{1}{\tau}\left(\frac{\rho^m+\rho^{m-1}}{2} v_h^m -\rho^{m-1}v^{m-1},w\right)\nonumber\\
  + a(\rho^{m-1}v^{m-1} + J^{m},v_h^m,w)
  + (2\eta^{m}Dv_h^m,Dw)- (\mbox{div} w,p_h^m)\nonumber\\
		+( \varphi^{m-1}\nabla \mu_h^m,w) 
		-(g\rho^{m},w) &= 0, \label{eq:S:1_NS}\\
   -(\mbox{div} v_h^m,q) &= 0, \label{eq:S:2_NS_2}\\
   \frac{1}{\tau}(\varphi_h^{m} - \varphi^{m-1},\Psi)
   -(\varphi^{m-1}v^{m-1},\nabla \Psi) + \frac{\tau}{\rho_{\min}}(|\varphi^{m-1}|^2 \nabla \mu_h^m,\nabla \Psi)
		+(\nabla \mu_h^m,\nabla \Psi) &= 0 \label{eq:S:3_CH1},\\
  \sigma\epsilon(\nabla \varphi_h^m,\nabla \Phi)+\frac{\sigma}{ \epsilon}(W_+^\prime(\varphi_h^m) + W_-^\prime(\varphi^{m-1}),\Phi)  
   - (\mu_h^m,\Phi) \nonumber\\
    +\left( \frac{S_\gamma}{2} (\varphi_h^m-\varphi^{m-1}) + \gamma_u^\prime(\varphi^{m-1}),
    \Phi\right)_{\partial\Omega} 
    &= 0,
    \label{eq:S:4_CH2}  
\end{align}
with
$J^{m} := -\frac{\partial \rho}{\partial\varphi}(\varphi^{m})\nabla \mu_h^{m}$,
$\rho^m := \rho(\varphi_h^m)$, $\rho^{m-1} := \rho(\varphi^{m-1})$, 
$\eta^{m} := \eta(\varphi_h^{m})$,
and $\eta^{m-1} := \eta(\varphi^{m-1})$.
The contact line energy $\gamma^\prime_u(\varphi^{m-1})$ is given by
$\gamma^\prime_u(\varphi^{m-1}) = \sigma_{lg}(\cos(\theta_{eq}) + B^mu)\vartheta^\prime(\varphi^{m-1})$,
with 
$B^mu:= \frac{1}{\tau}\int_{t_{m-1}}^{t_m} (Bu)(t)\dt$.

One can show, that \eqref{eq:S:1_NS}--\eqref{eq:S:4_CH2} admits a unique solution on every time instance, see \cite{BonKR-MCLOPT-ana}.
Especially, for every $u$ there exist unique sequences $(v_h^m,p_h^m,\varphi_h^m,\mu_h^m)_{m=1}^M$ that solve  \eqref{eq:S:1_NS}--\eqref{eq:S:4_CH2} for every $m = 1,\ldots,M$
and that is bounded by the initial data $v_0$ and $\varphi_0$.

Finally, for a sequence $ (\varphi_h^m)_{m=1}^M$ we introduce the function $\varphi_\tau$ as piecewiese constant with respect to time,
i.e. $\varphi_\tau(t) \equiv \varphi_h^m$
for all $t\in(t_{m-1},t_m)$.

Since $\varphi_\tau $ in fact is a fully discrete function, we can now state \cref{prob:O:P} as
\begin{equation}
 \tag{P$_h$}
 \label{prob:O:Ph}
 \min J(\varphi_\tau,u) = \sum_{m=1}^M \tau\omega^m \|\varphi_h^m - \varphi_d^m\|_{L^2(\Omega)}^2 + \frac{\alpha}{2} \sum_{m=1}^M \tau\|B^mu\|^2_{L^2(\partial\Omega)},
 \end{equation}
where $B^mu = \frac{1}{\tau}\int_{t_{m-1}}^{t_m} Bu(t)\dt$,  $\varphi_d^m = \frac{1}{\tau}\int_{t_{m-1}}^{t_m} \varphi_d(t)\dt$,  and $\omega^m = \frac{1}{\tau}\int_{t_{m-1}}^{t_m}\omega(t)\dt$ holds. 

\subsection{Numerical Implementation}
The adjoint and the gradient for \cref{prob:O:Ph} are derived using standard methods and follow directly from the discrete forward model and the functional, see~\cite{BonKR-MCLOPT-ana,GarHK_optContr_twoPhaseFlow}.
For details on the adjoint method for optimization with partial differential equations as constraints see~\cite{TroelOptiSteu,Hinze2009}.
We implement the solution scheme for the forward and adjoint model as well as the gradient in Python3 using FEniCS 2019.1.0~\cite{AlnaesBlechta2015a,fenics_book}.
For the solution of the arising linear systems and subsystems the software suite PETSc 3.8.4~\cite{petsc_webpage, petsc-user-ref, petsc-efficient} is applied.
The software IPOPT~\cite{Wachter2006} is applied for the solution of the optimization problem.

\section{Optimal Control of Droplets}
\label{sec:results}
In this section we demonstrate the applicability of our approach and framework.
At first, the general setup is described.
Then we describe the results for the design cases A and B as well as the active control cases C and D.
Finally, we extend the first example to sliding droplets.
Especially, for the pinning of the sliding droplet the most complex case D with its spatial and temporally varying contact angles is vital.

\subsection{General Setup}
\Cref{tab:parameters} states the physical and numerical parameters applied throughout this section.
Exemplarily, we used a water droplet with a diameter $l_d=\SI{5}{\milli\meter}$ surrounded by air on a solid surface with $\theta_{eq} = \SI{90}{\degree}$.
The time step was set to $\tau=\SI{0.001}{\second}$.
An adaptive mesh was utilized so that the interface between gas and liquid was resolved with at least four cells at all times ($h_{min} = \SI{0.04}{\milli\metre}$).

\begin{table}
\centering
\begin{tabular}{cccccccc}
	    \toprule
        $l_d/\si{\milli\metre}$&$\sigma_{gl}/\si{\milli\newton\per\meter}$&$\rho_l/\si{\kilo\gram\per\cubic\meter}$&$\eta_l/\si{\milli\Pa \s}$&$\rho_g/\si{\kilo\gram\per\cubic\meter}$&$\eta_g/\si{\milli\Pa \s}$&$g/\si{\meter\per\square\second}$&$\theta_{eq}/\si{\degree}$\\
	    \midrule
        5&72.86&	998		&1.0&1.2						&	0.018	&9.81  & 90 \\\addlinespace
        \midrule
        $\epsilon/\si{\milli\meter}$&$b/\si{\cubic\meter\per\joule \square\milli\meter\per\second}$&$\tau/\si{\second}$&$h_{min}/\si{\milli\meter}$ \\
        \midrule
        0.2&0.04&0.001&0.04&&&&\\
	    \bottomrule
\end{tabular}
        \caption{Physical and numerical parameters of droplet (water), surrounding gas (air) and solid surface.}
\label{tab:parameters}
\end{table}

The desired shapes $\varphi_d$ were created by simulating the forward model \cref{eq:S:1_NS}--\cref{eq:S:4_CH2} from a suitable initial value, which is specified for the individual setups, until the equilibrium droplet shape was reached.

The values of $R$ and $S$ for the four control cases described in~\cref{sec:problems} are given in~\cref{tab:cases} for the two numerical examples presented in \Cref{sec:sym} and \Cref{sec:pin}.
In both examples the stripes as well as the switching points were equally distributed over the length of the bottom wall and the time horizon of $I=[0.0,T]$%
with $T=\SI{1}{\second}$, respectively, compare \Cref{ssec:control}.

The weight $\omega$ was chosen as 
\begin{align*}
        \omega(t) = \begin{cases}
				1 & \mbox{ if } t \in (0,0.8)\si{\second},\\
                        10^5 & \mbox{ if } t \in [0.8,1.0]\si{\second},
                    \end{cases}
\end{align*}
to promote a good matching of $\varphi$ and $\varphi_d$ at the end of the optimization horizon between $\SI{0.8}{\second}$ and $\SI{1.0}{\second}$.
We stopped the optimizer as soon as the initial gradient $\nabla J$ was reduced by a factor of $\frac{1}{100}$.

\begin{table}
\centering
\begin{tabular}{cccccc}
	    \toprule
        &\multicolumn{4}{c}{Prob. 1}&Prob. 2\\
        \cmidrule(r){2-5}\cmidrule(r){6-6}
        &A&B&C&D&D\\
        \cmidrule(r){2-5}\cmidrule(r){6-6}
        R&1&1&5&5&10\\
        S&1&5&1&5&10\\
        \bottomrule
\end{tabular}
        \caption{Values of $R$ and $S$ for the control cases A to D for both problems described in~\cref{sec:problems}.}
\label{tab:cases}
\end{table}

\subsection{Droplet on a Horizontal Surface}\label{sec:sym}
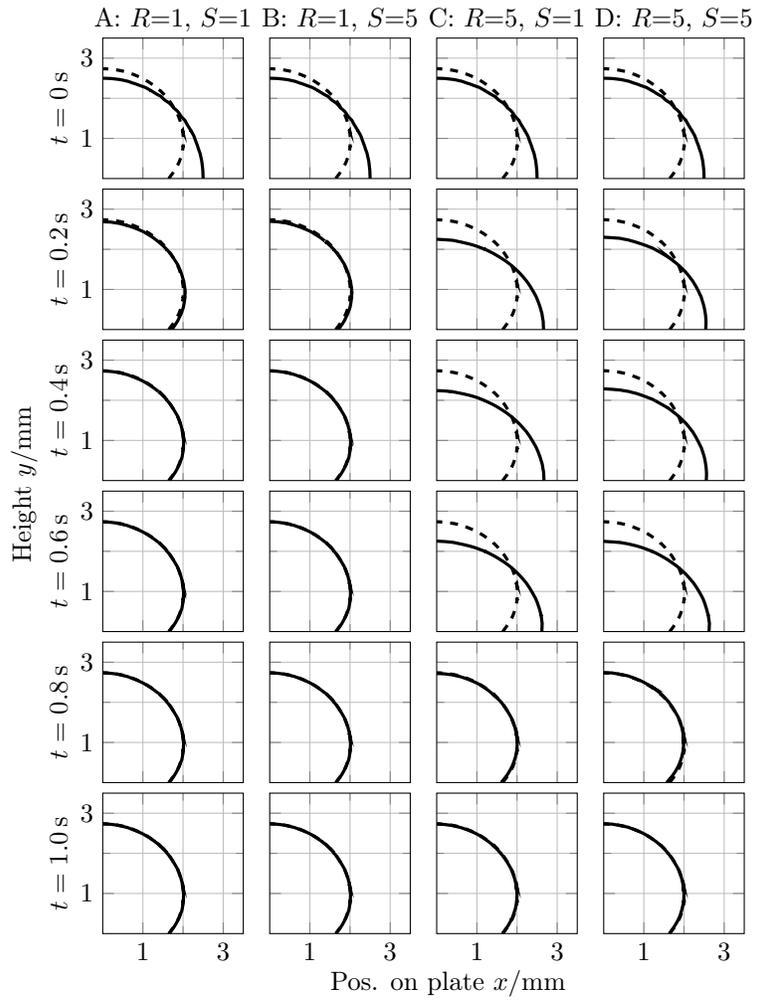
\begin{figure}[h]
\tikzsetnextfilename{sym_isolines}
\centering
\begin{tikzpicture}
	\centering
	\begin{groupplot}[
			group style = {
				group size = 4 by 6
				,vertical sep=4pt,horizontal sep=10pt
				,ylabels at=edge left
				,xlabels at=edge bottom
               	,yticklabels at=edge left 
				,xticklabels at=edge bottom
			},
    		width = 4.0cm%
			,xlabel near ticks
			,ylabel near ticks
			,unit vector ratio=1 1 1
			, xmin=0, xmax=3.5
			, ymin=0, ymax=3.5
			, ytick={1,3}
			, xtick={1,3}
			, grid=both
			, minor tick num=1
			, ylabel shift = -2 pt
			, title style={yshift=-7pt}
			,x filter/.code={\pgfmathparse{#1*1000}\pgfmathresult},
		    ,y filter/.code={\pgfmathparse{#1*1000}\pgfmathresult}
			]
			\nextgroupplot[title={A: $R{=}1$, $S{=}1$}, ylabel={$t=\SI{0}{\second}$}] 
				\addplot[very thick, smooth] plot file{data/symmetric/forward_opt_x1_t1/isolines/isoline_phi0_tau0.0.dat};
				\addplot[very thick, smooth, dashed] plot file{data/symmetric/isoline_phid.dat};
			\nextgroupplot[title={B: $R{=}1$, $S{=}5$}] 
				\addplot[very thick, smooth] plot file{data/symmetric/forward_opt_x5_t1/isolines/isoline_phi0_tau0.0.dat};
				\addplot[very thick, smooth, dashed] plot file{data/symmetric/isoline_phid.dat};
			\nextgroupplot[title={C: $R{=}5$, $S{=}1$}] 
				\addplot[very thick, smooth] plot file{data/symmetric/forward_opt_x1_t5/isolines/isoline_phi0_tau0.0.dat};
				\addplot[very thick, smooth, dashed] plot file{data/symmetric/isoline_phid.dat};
			\nextgroupplot[title={D: $R{=}5$, $S{=}5$}] 
				\addplot[very thick, smooth] plot file{data/symmetric/forward_opt_x5_t5/isolines/isoline_phi0_tau0.0.dat};
				\addplot[very thick, smooth, dashed] plot file{data/symmetric/isoline_phid.dat};
			\nextgroupplot[ylabel={$t=\SI{0.2}{\second}$}] 
				\addplot[very thick, smooth] plot file{data/symmetric/forward_opt_x1_t1/isolines/isoline_phi0_tau0.2.dat};
				\addplot[very thick, smooth, dashed] plot file{data/symmetric/isoline_phid.dat};
			\nextgroupplot[] 
				\addplot[very thick, smooth] plot file{data/symmetric/forward_opt_x5_t1/isolines/isoline_phi0_tau0.2.dat};
				\addplot[very thick, smooth, dashed] plot file{data/symmetric/isoline_phid.dat};
			\nextgroupplot[] 
				\addplot[very thick, smooth] plot file{data/symmetric/forward_opt_x1_t5/isolines/isoline_phi0_tau0.2.dat};
				\addplot[very thick, smooth, dashed] plot file{data/symmetric/isoline_phid.dat};
			\nextgroupplot[] 
				\addplot[very thick, smooth] plot file{data/symmetric/forward_opt_x5_t5/isolines/isoline_phi0_tau0.2.dat};
				\addplot[very thick, smooth, dashed] plot file{data/symmetric/isoline_phid.dat};
			\nextgroupplot[ylabel={$t=\SI{0.4}{\second}$}] 
				\addplot[very thick, smooth] plot file{data/symmetric/forward_opt_x1_t1/isolines/isoline_phi0_tau0.4.dat};
				\addplot[very thick, smooth, dashed] plot file{data/symmetric/isoline_phid.dat};
			\nextgroupplot[] 
				\addplot[very thick, smooth] plot file{data/symmetric/forward_opt_x5_t1/isolines/isoline_phi0_tau0.4.dat};
				\addplot[very thick, smooth, dashed] plot file{data/symmetric/isoline_phid.dat};
			\nextgroupplot[] 
				\addplot[very thick, smooth] plot file{data/symmetric/forward_opt_x1_t5/isolines/isoline_phi0_tau0.4.dat};
				\addplot[very thick, smooth, dashed] plot file{data/symmetric/isoline_phid.dat};
			\nextgroupplot[] 
				\addplot[very thick, smooth] plot file{data/symmetric/forward_opt_x5_t5/isolines/isoline_phi0_tau0.4.dat};
				\addplot[very thick, smooth, dashed] plot file{data/symmetric/isoline_phid.dat};
				\nextgroupplot[ylabel={$t=\SI{0.6}{\second}$}] 
				\addplot[very thick, smooth] plot file{data/symmetric/forward_opt_x1_t1/isolines/isoline_phi0_tau0.6.dat};
				\addplot[very thick, smooth, dashed] plot file{data/symmetric/isoline_phid.dat};
			\nextgroupplot[] 
				\addplot[very thick, smooth] plot file{data/symmetric/forward_opt_x5_t1/isolines/isoline_phi0_tau0.6.dat};
				\addplot[very thick, smooth, dashed] plot file{data/symmetric/isoline_phid.dat};
			\nextgroupplot[] 
				\addplot[very thick, smooth] plot file{data/symmetric/forward_opt_x1_t5/isolines/isoline_phi0_tau0.6.dat};
				\addplot[very thick, smooth, dashed] plot file{data/symmetric/isoline_phid.dat};
			\nextgroupplot[] 
				\addplot[very thick, smooth] plot file{data/symmetric/forward_opt_x5_t5/isolines/isoline_phi0_tau0.6.dat};
				\addplot[very thick, smooth, dashed] plot file{data/symmetric/isoline_phid.dat};
			\nextgroupplot[ylabel={$t=\SI{0.8}{\second}$}] 
				\addplot[very thick, smooth] plot file{data/symmetric/forward_opt_x1_t1/isolines/isoline_phi0_tau0.8.dat};
				\addplot[very thick, smooth, dashed] plot file{data/symmetric/isoline_phid.dat};
			\nextgroupplot[] 
				\addplot[very thick, smooth] plot file{data/symmetric/forward_opt_x5_t1/isolines/isoline_phi0_tau0.8.dat};
				\addplot[very thick, smooth, dashed] plot file{data/symmetric/isoline_phid.dat};
			\nextgroupplot[] 
				\addplot[very thick, smooth] plot file{data/symmetric/forward_opt_x1_t5/isolines/isoline_phi0_tau0.8.dat};
				\addplot[very thick, smooth, dashed] plot file{data/symmetric/isoline_phid.dat};
			\nextgroupplot[] 
				\addplot[very thick, smooth] plot file{data/symmetric/forward_opt_x5_t5/isolines/isoline_phi0_tau0.8.dat};
				\addplot[very thick, smooth, dashed] plot file{data/symmetric/isoline_phid.dat};
			\nextgroupplot[ylabel={$t=\SI{1.0}{\second}$}] 
				\addplot[very thick, smooth] plot file{data/symmetric/forward_opt_x1_t1/isolines/isoline_phi0_tau1.0.dat};
				\addplot[very thick, smooth, dashed] plot file{data/symmetric/isoline_phid.dat};
			\nextgroupplot[] 
				\addplot[very thick, smooth] plot file{data/symmetric/forward_opt_x5_t1/isolines/isoline_phi0_tau1.0.dat};
				\addplot[very thick, smooth, dashed] plot file{data/symmetric/isoline_phid.dat};
			\nextgroupplot[] 
				\addplot[very thick, smooth] plot file{data/symmetric/forward_opt_x1_t5/isolines/isoline_phi0_tau1.0.dat};
				\addplot[very thick, smooth, dashed] plot file{data/symmetric/isoline_phid.dat};
			\nextgroupplot[] 
				\addplot[very thick, smooth] plot file{data/symmetric/forward_opt_x5_t5/isolines/isoline_phi0_tau1.0.dat};		%
				\addplot[very thick, smooth, dashed] plot file{data/symmetric/isoline_phid.dat};
		\end{groupplot}
		\node[above, rotate=90] at ($(current bounding box.west) + (2pt,0)$) {Height $y/\si{\milli\meter}$};
		\node[below] at ($(current bounding box.south) + (25pt,3pt)$) {Pos. on plate $x/\si{\milli\meter}$};
	\end{tikzpicture}
	\caption{Development of the receding droplets over time for the optimization cases A to D. Due to the symmetric droplet only the right half of the isoline for $\varphi=0$ is displayed. The desired shape and position of the droplet $\varphi_d$ is included as the dashed line.}	
	\label{fig:sym_isolines}
\end{figure}

At first, we report on the optimal control of the first example described in~\cref{sec:problems}.
The domain was of dimension $\Omega = l_d\times l_d$ and we made use of the symmetrical droplet.
In this way, we assumed symmetric conditions at the left side of the domain and solid walls at the top, bottom and right side of the domain.
The bottom wall contained the control of the contact angle.
The desired shape $\varphi_d$ with an equilibrium contact angle of $\SI{135}{\degree}$ is shown as the dashed line in~\cref{fig:sym_isolines}.
In the first row of~\cref{fig:sym_isolines} the initial droplets with $Bu=0$ and $\theta_{eq}=\SI{90}{\degree}$ are shown.
The values for $R$ and $S$ for the control cases described in~\cref{sec:problems} are given in~\cref{tab:cases}.
In this example, at most five stripes with each five switching points in time were considered.

We discuss the following aspects of our numerical results:

\paragraph{Resulting Droplet Shapes}
The droplet shapes calculated with the resulting optimal control values $Bu$ for the cases A to D are depicted in~\cref{fig:sym_isolines}.
The solid lines represent the isolines where $\varphi=0$.
We show the shape and position of the droplet at five instances in time.
Note, that only the relevant clipping of $\Omega$ is displayed in~\cref{fig:sym_isolines}.

\paragraph{Relative Mismatch}
The relative mismatch between the actual droplet and the desired shape over time,
\begin{align}
		\frac{\Delta\varphi(t)}{\Delta\varphi^0}=\sqrt{\frac{\int_\Omega (\varphi(t)-\varphi_d)^2\dx}{\int_\Omega (\varphi^0-\varphi_d)^2\dx}}\;,
\end{align}
which is normalized with the mismatch at $t=\SI{0.0}{\second}$, is displayed in the top part of~\cref{fig:sym_pos_diff_strength}.

\paragraph{Optimal Controls}
The corresponding optimal controls $Bu$ for the cases A to D are plotted in~\cref{fig:sym_controls}.
In the first two columns the controls are constant in time, whereas the controls are constant in space in the first and third column.
The dashed lines represent the control which was used to create the desired shape $\varphi_d$ 
(i.e., $\cos(\SI{135}{\degree}) \approx -0.7071$).

\paragraph{Relative Controller Strength}
The relative controller strength per time interval $r$ is calculated with
\begin{align}
		\frac{E_r}{E_\text{A}}  
		=\frac{\frac{1}{S}\sqrt{\sum_{s=1}^S u_{rs}^2}}{|u_A|}\;.
\end{align}
The control strength per time interval is normalized to the (constant) control strength needed in case A. We further assume, that
the control costs scale linearly with the length of the control domain, which leads to the normalization with 
$\frac{1}{S}$ in the numerator.
It is plotted in the lower part of~\cref{fig:sym_pos_diff_strength}.

\subsubsection{Cases A and B}
Comparing the design cases A and B (first and second column in~\cref{fig:sym_isolines} it is evident that the droplet behave almost identical: the shapes and the positions of the contact lines are very similar. 
The desired shape is reached with very high accuracy after a short time (see $t=\SI{0.4}{\second}$) in both cases.
The mismatches $\Delta \varphi$ displayed in the top part of~\cref{fig:sym_pos_diff_strength} (solid lines with triangular and circular markers) decline with the same slope over time and reach almost zero at around $t=\SI{0.4}{\second}$.

Now we focus on the first and second column of~\cref{fig:sym_controls} to include the optimal controls into our discussion.
In case A we only allowed for a single control action for the whole surface and time horizon.
Subsequently, only one value is displayed in column one.
As expected from the shape of the droplet, the control value is very close to the value which was used to create the desired droplet $\varphi_d$, see the dashed line in the lower left panel of~\cref{fig:sym_controls}.

In case B the stripe between \SI{2}{\milli\meter} and \SI{3}{\milli\meter} has a control value which is slightly lower than the equilibrium value obtained again in the stripe $[1,2]\si{\milli\meter}$.
This results in a more rapid receding of the droplet on that stripe in case B.
However, as the droplet spends only little time on that stripe the influence on the overall dynamics of the droplet are low.
Since in case B only a portion of the bottom controls are active at any time, the controller strength is significantly lower compared to case A, see the lower part of~\cref{fig:sym_pos_diff_strength}.

\subsubsection{Cases C and D}
\begin{figure}[h]
\tikzsetnextfilename{sym_pos_diff_strength}
\centering
\begin{tikzpicture}[spy using outlines=
		{circle, magnification=2, connect spies}]
	\centering
	\begin{groupplot}[
			group style = {
				group size = 1 by 2
				,vertical sep=12pt,horizontal sep=12pt
				,xlabels at=edge bottom
				,xticklabels at=edge bottom
			},
    		width = 7cm, height = 5cm
			,xlabel near ticks
			, xmin=0, xmax=1
			, legend pos=north east
		    , legend columns=2
			, legend style={font=\small}
			, xlabel shift = -3 pt
            , xlabel={time $t/\si{\second}$}
			, ylabel style={at={(-0.15,0.5)},align=center, font=\small\linespread{1.0}\selectfont}
            , xlabel style={align=center, font=\small\linespread{1.0}\selectfont}
		,ymode=log,grid=major
		, ymin=1e-3,ymax=2
,log ticks with fixed point
			]
            \nextgroupplot[ylabel={rel. mismatch\\$\Delta\varphi/\Delta\varphi^0$}, legend to name=bla,ymode=log, ymin=1e-2,ymax=2
        ,] 
			    \addplot[thick,mark=triangle,mark repeat=80,mark phase=40] table [x index=0, y expr=sqrt(\thisrowno{1}/1.4087e-6)] {data/symmetric/forward_opt_x1_t1/difference.dat};
				\addplot[thick,mark=*,mark repeat=80] table [x index=0, y expr=sqrt(\thisrowno{1}/1.4087e-6)] {data/symmetric/forward_opt_x5_t1/difference.dat};	
				\addplot[thick,mark=square,mark repeat=80,mark phase=40] table [x index=0, y expr=sqrt(\thisrowno{1}/1.4087e-6)] {data/symmetric/forward_opt_x1_t5/difference.dat};
				\addplot[thick,mark=o,mark repeat=80] table [x index=0, y expr=sqrt(\thisrowno{1}/1.4087e-6)] {data/symmetric/forward_opt_x5_t5/difference.dat};
				
                \addlegendentry{A: $R=1, S=1$}
				\addlegendentry{B: $R=1, S=5$}
				\addlegendentry{C: $R=5, S=1$}
				\addlegendentry{D: $R=5, S=5$}
				\coordinate (left) at (rel axis cs:0,0);
				\coordinate (right) at (rel axis cs:1,0);
				\coordinate (spypoint) at (axis cs:0.15,0.25);
				\coordinate (magnifyglass) at (axis cs:0.4,0.55);
				\nextgroupplot[ylabel={rel. control strength\\$E_r/E_\text{A}$}
        ] 
		\addplot[thick, const plot mark mid,mark=triangle] table [x expr=\thisrowno{0}+0.01, y expr=(\thisrowno{1})] {data/symmetric/forward_opt_x1_t1/strength.dat};
		\addplot[thick, const plot mark mid,mark=*] table [x expr=\thisrowno{0}-0.01, y expr=(\thisrowno{1})] {data/symmetric/forward_opt_x5_t1/strength.dat};
				\addplot[thick, const plot mark mid,mark=square] table [x expr=\thisrowno{0}-0.01, y expr=(\thisrowno{1})] {data/symmetric/forward_opt_x1_t5/strength.dat};
				\addplot[thick, const plot mark mid,mark=o] table [x expr=\thisrowno{0}+0.01, y expr=(\thisrowno{1})] {data/symmetric/forward_opt_x5_t5/controls/strength.dat};
				\coordinate (top) at (rel axis cs:0,1);
				\coordinate (bot) at (rel axis cs:1,0);
				\node[anchor=south west,align=left,font=\scriptsize\linespread{0.8}\selectfont,fill=white,inner sep=1pt] at (axis cs:0.61,0.002) {intervals shifted\\for visibility};
		\end{groupplot}
			\path (top)--(bot) coordinate[midway] (group center);
			\path (left)--(right) coordinate[midway] (group center2);
            \node [right=8mm,anchor=center,rotate=90] at ($(group c1r1.south east)!0.5!(group c1r2.north east)$) {\pgfplotslegendfromname{bla}};

\end{tikzpicture}
\caption{Development of the relative mismatch (top) and the relative control strength for the cases A, B, C and D.}
\label{fig:sym_pos_diff_strength}
\end{figure}
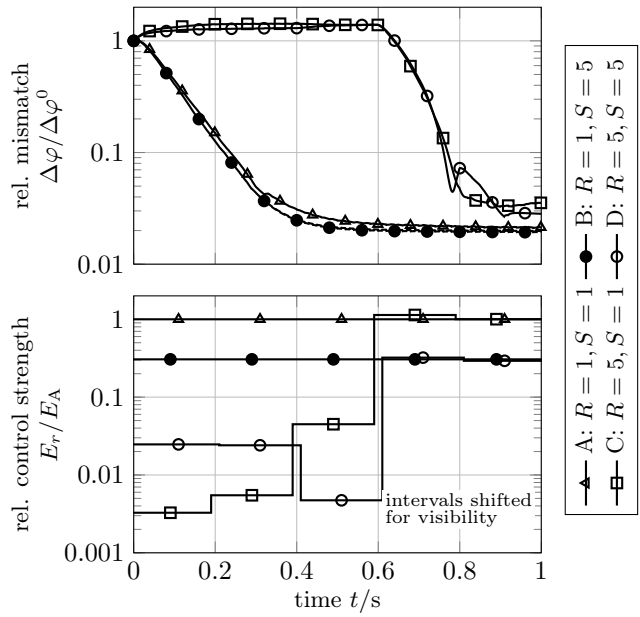

From comparing the shape of the droplets in the active cases C and D (third and fourth column in~\cref{fig:sym_isolines} we again notice that the droplets behave almost identical: the shapes and the positions of the contact lines are very similar. 
The desired shape is reached relatively late but still a very high accuracy after $t=\SI{0.8}{\second}$ in both cases.
This stems from the weight $\omega$ defined above, which was strongly penalizing the interval $[0.8,1.0]$.
Before \SI{0.6}{\second} the droplets do not move at all.
This can be seen by looking at the mismatches of $\varphi$ displayed in the top part of~\cref{fig:sym_pos_diff_strength} (solid lines with rectangular and open circular markers).
Until \SI{0.6}{\second} the mismatch gets even higher than the initial mismatch. 
This is due to the initial droplet adapting to gravity. 
The subsequent decline however is very rapid and happens with almost the same slope over time for the cases C and D.
Finally, it reaches a small value at around $t=\SI{0.8}{\second}$.

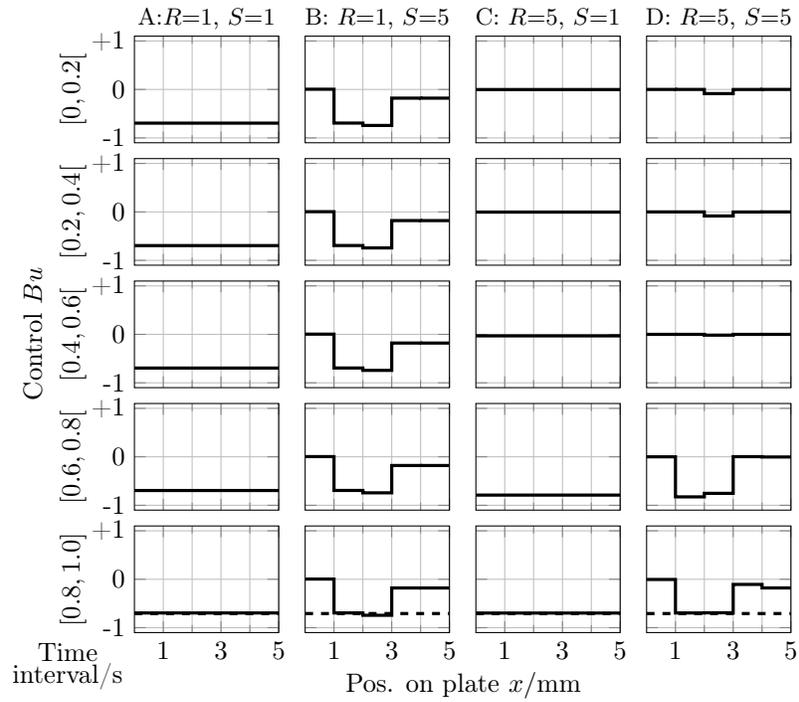
\begin{figure}[h]
\tikzsetnextfilename{sym_controls}
\centering
\begin{tikzpicture}
	\centering
	\begin{groupplot}[
			group style = {
				group size = 4 by 5
				,vertical sep=6pt,horizontal sep=10pt
				,ylabels at=edge left
				,xlabels at=edge bottom
               	,yticklabels at=edge left 
				,xticklabels at=edge bottom
			},
    		width = 3.5cm, height = 3cm
			,xlabel near ticks
			,ylabel near ticks
			, xmin=0, xmax=5
			, ymin=-1.1, ymax=1.1
			, ytick={-1,0,+1}
			, yticklabels={-1, 0, +1}
			, xtick={1,3,5}
			, grid=both
			, minor x tick num=1
			, ylabel shift = -5 pt
			, title style={yshift=-7pt, font=\small\linespread{0.8}\selectfont}
		    ,x filter/.code={\pgfmathparse{#1*1000}\pgfmathresult},
			]
			\nextgroupplot[title={A:$R{=}1$, $S{=}1$}, ylabel={$[0,0.2[$}] 
				\addplot[mark=none, very thick, black ] coordinates {(0,-6.953603509534129756e-01) (0.005,-6.953603509534129756e-01)};
			\nextgroupplot[title={B: $R{=}1$, $S{=}5$}] 
				\addplot[very thick, const plot mark left]  table [x index=0, y index=1]{data/symmetric/forward_opt_x5_t1/control_opt.dat};
			\nextgroupplot[title={C: $R{=}5$, $S{=}1$}] 
				\addplot[mark=none, very thick, black ] coordinates {(0,-2.274784962014267907e-03) (0.005,-2.274784962014267907e-03)};
			\nextgroupplot[title={D: $R{=}5$, $S{=}5$}] 
				\addplot[very thick, const plot mark left]  table [x index=0, y index=1]{data/symmetric/forward_opt_x5_t5/controls/control_opt_0.dat};
			\nextgroupplot[ylabel={$[0.2,0.4[$}] 
					\addplot[mark=none, very thick, black ] coordinates {(0,-6.953603509534129756e-01) (0.005,-6.953603509534129756e-01)};
			\nextgroupplot[] 
				\addplot[very thick, const plot mark left]  table [x index=0, y index=1]{data/symmetric/forward_opt_x5_t1/control_opt.dat};
			\nextgroupplot[] 
				\addplot[mark=none, very thick, black ] coordinates {(0,-3.814516658079286811e-03) (0.005,-3.814516658079286811e-03)};
			\nextgroupplot[] 
				\addplot[very thick, const plot mark left]  table [x index=0, y index=1]{data/symmetric/forward_opt_x5_t5/controls/control_opt_1.dat};
			\nextgroupplot[ylabel={$[0.4,0.6[$}] 
	\addplot[mark=none, very thick, black ] coordinates {(0,-6.953603509534129756e-01) (0.005,-6.953603509534129756e-01)};

			\nextgroupplot[] 
				\addplot[very thick, const plot mark left]  table [x index=0, y index=1]{data/symmetric/forward_opt_x5_t1/control_opt.dat};
			\nextgroupplot[] 
				\addplot[mark=none, very thick, black ] coordinates {(0,-3.120251671702286428e-02) (0.005,-3.120251671702286428e-02)};
			\nextgroupplot[] 
				\addplot[very thick, const plot mark left]  table [x index=0, y index=1]{data/symmetric/forward_opt_x5_t5/controls/control_opt_2.dat};
			\nextgroupplot[ylabel={$[0.6,0.8[$}] 
	\addplot[mark=none, very thick, black ] coordinates {(0,-6.953603509534129756e-01) (0.005,-6.953603509534129756e-01)};

			\nextgroupplot[] 
				\addplot[very thick, const plot mark left]  table [x index=0, y index=1]{data/symmetric/forward_opt_x5_t1/control_opt.dat};
			\nextgroupplot[] 
				\addplot[mark=none, very thick, black ] coordinates {(0,-7.898936583693039193e-01) (0.005,-7.898936583693039193e-01)};
			\nextgroupplot[] 
				\addplot[very thick, const plot mark left]  table [x index=0, y index=1]{data/symmetric/forward_opt_x5_t5/controls/control_opt_3.dat};
			\nextgroupplot[ylabel={$[0.8,1.0]$}] 
				\addplot[mark=none, very thick, black ] coordinates {(0,-6.953603509534129756e-01) (0.005,-6.953603509534129756e-01)};
				\addplot[mark=none, very thick, black, dashed] coordinates {(0,-7.071e-01) (0.005,-7.071e-01)};
			\nextgroupplot[] 
				\addplot[very thick, const plot mark left]  table [x index=0, y index=1]{data/symmetric/forward_opt_x5_t1/control_opt.dat};
				\addplot[mark=none, very thick, black, dashed] coordinates {(0,-7.071e-01) (0.005,-7.071e-01)};
			\nextgroupplot[] 
				\addplot[mark=none, very thick, black ] coordinates {(0,-6.975643461268303813e-01) (0.005,-6.975643461268303813e-01)};
				\addplot[mark=none, very thick, black, dashed] coordinates {(0,-7.071e-01) (0.005,-7.071e-01)};
			\nextgroupplot[] 
				\addplot[very thick, const plot mark left]  table [x index=0, y index=1]{data/symmetric/forward_opt_x5_t5/controls/control_opt_4.dat};
				\addplot[mark=none, very thick, black, dashed] coordinates {(0,-7.071e-01) (0.005,-7.071e-01)};
		\end{groupplot}
		\node[above, rotate=90] at ($(current bounding box.west) + (-0pt,0)$) {Control $Bu$};
		\node[below] at ($(current bounding box.south) + (20pt,2pt)$) {Pos. on plate $x/\si{\milli\meter}$};
		\node[above,align=center, execute at begin node=\setlength{\baselineskip}{0pt}] at ($(current bounding box.south west) + (+20pt,+3pt)$) {Time\\interval/s};
	\end{tikzpicture}
	\caption{Development of the controls over time for the optimization cases A to D. Due to the symmetric droplet only the right half of the controls is displayed. The control which was used to create the desired droplet $\varphi_d$ is displayed as the dashed line at the bottom row.}	
	\label{fig:sym_controls}
\end{figure}

By investigating the controls in the third and fourth column of~\cref{fig:sym_controls} the explanation for the observed droplet shapes becomes obvious.
In the first three intervals the control action is approximately zero.
Only for the last intervals control actions are needed to reach the desired shape in time.
This leads to very low relative controller strengths required to reach the optimization target, see the lower panel of~\cref{fig:sym_pos_diff_strength}.
The most complex control sequence is calculated in case D.
Here, the droplet moves in a very versatile and controlled fashion until it reaches the desired shape with a very low relative controller strength.

\bigskip

For all cases A to D we were able to calculate optimal controls to reach the desired shape for a specific time interval: the shapes of the droplets match the desired shape in all cases almost perfectly (\cref{fig:sym_isolines} and the mismatch gets very low (\cref{fig:sym_pos_diff_strength}, top).
In cases C and D, the mismatch decreases slower than in cases A and B. 
This indicates, that a shorter time horizon would be sufficient for the control task.
In all cases the final control values for the stripe between \SI{1}{\milli\meter} and \SI{2}{\milli\meter} is almost the same as the one used for the desired shape (\cref{fig:sym_controls}). 
This is expected, since the desired shape is stable situation with a constant contact angle. Moreover, the final contact point is located in this stripe, thus defining the shape of the droplet.
Subsequently, the required relative controller strengths in the cases C and D are much smaller than in the cases A and B (\cref{fig:sym_pos_diff_strength}, bottom).
This is due to the fact, that the more complex controls can act more locally precise with respect to space and time. Thus they are not active (i.e. $Bu=0$) in large parts of the spatial and temporal domain.

\subsection{Pinning a Sliding Droplet} \label{sec:pin}
To demonstrate the general applicability of our approach we extend the example discussed in~\cref{sec:sym} to the optimal pinning of a sliding droplet.
The pinning of a sliding droplet at a specific position on a solid surface while maintaining a desired shape can have interesting implications for technical applications.
At the same time, this is a challenging task.

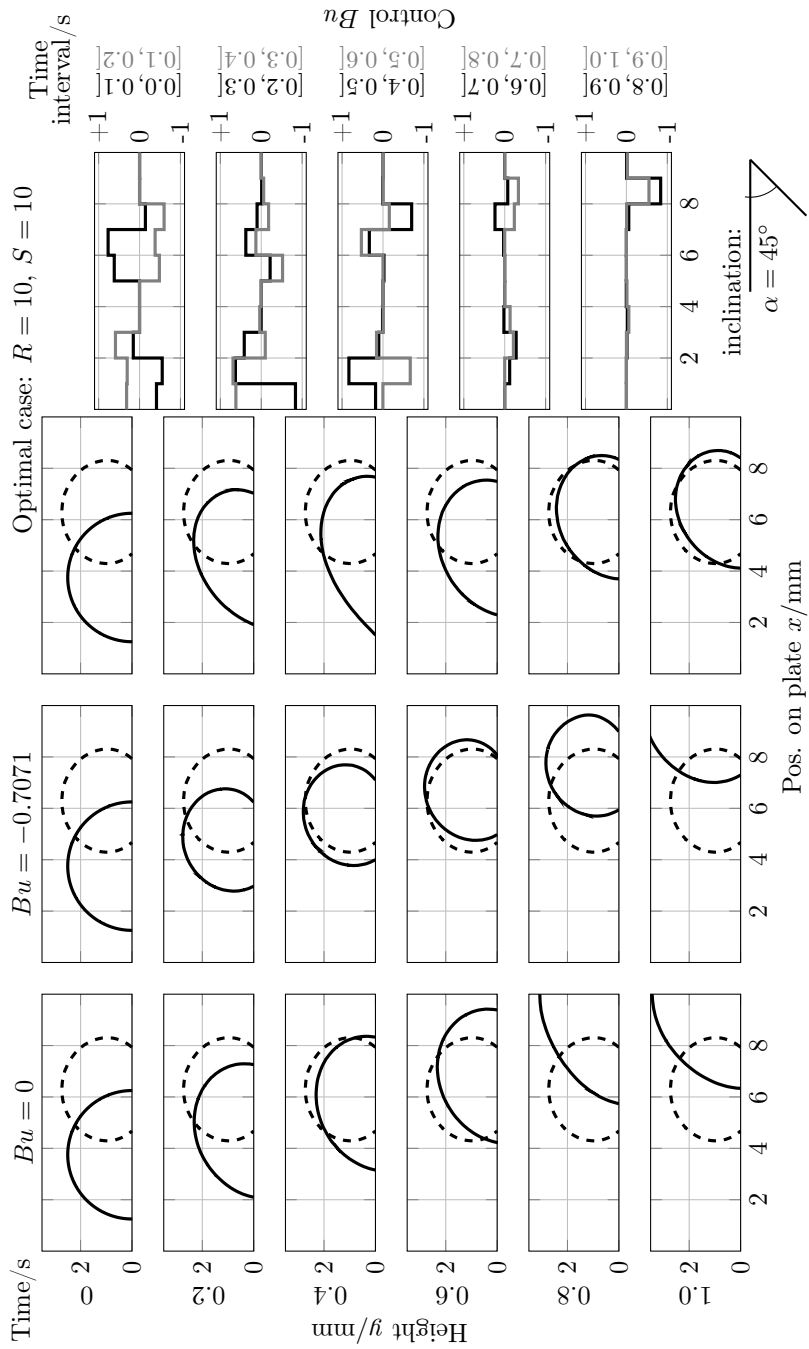
\begin{sidewaysfigure}[h]
\tikzsetnextfilename{sli_results}
\centering
\begin{tikzpicture}
	\centering
	\begin{groupplot}[
			group style = {
				group size = 3 by 6
				,vertical sep=12pt,horizontal sep=12pt
				,ylabels at=edge left
				,xlabels at=edge bottom
               	,yticklabels at=edge left 
				,xticklabels at=edge bottom
				, group name=left plots
			},
    		width = 5cm%
			,xlabel near ticks
			,ylabel near ticks
			,unit vector ratio=1 1 1
			, xmin=0, xmax=10
			, ymin=0, ymax=3.5
			, ytick={0, 2}
			, xtick={2,4,6,8}
			, grid=both
			, ylabel shift = -2 pt
			, title style={yshift=-6pt}
			, yticklabel style={
						/pgf/number format/fixed,
						/pgf/number format/precision=5
			}
			,scaled y ticks=false
			,x filter/.code={\pgfmathparse{#1*1000}\pgfmathresult},
		    ,y filter/.code={\pgfmathparse{#1*1000}\pgfmathresult}
			]
			\nextgroupplot[title={$Bu=0$}, ylabel={$0$}] 
				\addplot[very thick, smooth] plot file{data/sliding/forward_zero/isolines/isoline_phi0_tau0.0.dat};
				\addplot[very thick, smooth, dashed] plot file{data/sliding/isoline_phid.dat};
			\nextgroupplot[title={$Bu=-0.7071$}] 
				\addplot[very thick, smooth] plot file{data/sliding/forward_135/isolines/isoline_phi0_tau0.0.dat};
				\addplot[very thick, smooth, dashed] plot file{data/sliding/isoline_phid.dat};
			\nextgroupplot[] 
				\addplot[very thick, smooth] plot file{data/sliding/forward_opt_x10_t10/isolines/isoline_phi0_tau0.0.dat};
				\addplot[very thick, smooth, dashed] plot file{data/sliding/isoline_phid.dat};
			\nextgroupplot[ylabel={$0.2$}] 
				\addplot[very thick, smooth] plot file{data/sliding/forward_zero/isolines/isoline_phi0_tau0.2.dat};
				\addplot[very thick, smooth, dashed] plot file{data/sliding/isoline_phid.dat};
			\nextgroupplot[] 
				\addplot[very thick, smooth] plot file{data/sliding/forward_135/isolines/isoline_phi0_tau0.2.dat};
				\addplot[very thick, smooth, dashed] plot file{data/sliding/isoline_phid.dat};
			\nextgroupplot[] 
				\addplot[very thick, smooth] plot file{data/sliding/forward_opt_x10_t10/isolines/isoline_phi0_tau0.2.dat};
				\addplot[very thick, smooth, dashed] plot file{data/sliding/isoline_phid.dat};
			\nextgroupplot[ylabel={$0.4$}] 
				\addplot[very thick, smooth] plot file{data/sliding/forward_zero/isolines/isoline_phi0_tau0.4.dat};
				\addplot[very thick, smooth, dashed] plot file{data/sliding/isoline_phid.dat};
			\nextgroupplot[] 
				\addplot[very thick, smooth] plot file{data/sliding/forward_135/isolines/isoline_phi0_tau0.4.dat};
				\addplot[very thick, smooth, dashed] plot file{data/sliding/isoline_phid.dat};
			\nextgroupplot[] 
				\addplot[very thick, smooth] plot file{data/sliding/forward_opt_x10_t10/isolines/isoline_phi0_tau0.4.dat};
				\addplot[very thick, smooth, dashed] plot file{data/sliding/isoline_phid.dat};
			\nextgroupplot[ylabel={$0.6$}] 
				\addplot[very thick, smooth] plot file{data/sliding/forward_zero/isolines/isoline_phi0_tau0.6.dat};
				\addplot[very thick, smooth, dashed] plot file{data/sliding/isoline_phid.dat};
			\nextgroupplot[] 
				\addplot[very thick, smooth] plot file{data/sliding/forward_135/isolines/isoline_phi0_tau0.6.dat};
				\addplot[very thick, smooth, dashed] plot file{data/sliding/isoline_phid.dat};
			\nextgroupplot[] 
				\addplot[very thick, smooth] plot file{data/sliding/forward_opt_x10_t10/isolines/isoline_phi0_tau0.6.dat};
				\addplot[very thick, smooth, dashed] plot file{data/sliding/isoline_phid.dat};
			\nextgroupplot[ylabel={$0.8$}] 
				\addplot[very thick, smooth] plot file{data/sliding/forward_zero/isolines/isoline_phi0_tau0.8.dat};
				\addplot[very thick, smooth, dashed] plot file{data/sliding/isoline_phid.dat};
			\nextgroupplot[] 
				\addplot[very thick, smooth] plot file{data/sliding/forward_135/isolines/isoline_phi0_tau0.8.dat};
				\addplot[very thick, smooth, dashed] plot file{data/sliding/isoline_phid.dat};
			\nextgroupplot[] 
				\addplot[very thick, smooth] plot file{data/sliding/forward_opt_x10_t10/isolines/isoline_phi0_tau0.8.dat};
				\addplot[very thick, smooth, dashed] plot file{data/sliding/isoline_phid.dat};
			\nextgroupplot[ylabel={$1.0$}] 
				\addplot[very thick, smooth] plot file{data/sliding/forward_zero/isolines/isoline_phi0_tau1.0.dat};
				\addplot[very thick, smooth, dashed] plot file{data/sliding/isoline_phid.dat};
			\nextgroupplot[] 
				\addplot[very thick, smooth] plot file{data/sliding/forward_135/isolines/isoline_phi0_tau1.0.dat};
				\addplot[very thick, smooth, dashed] plot file{data/sliding/isoline_phid.dat};
			\nextgroupplot[] 
				\addplot[very thick, smooth] plot file{data/sliding/forward_opt_x10_t10/isolines/isoline_phi0_tau1.0.dat};
				\addplot[very thick, smooth, dashed] plot file{data/sliding/isoline_phid.dat};
		\end{groupplot}
		\begin{groupplot}[
			group style = {
				,group size = 1 by 5
				,vertical sep=12pt,horizontal sep=12pt
				,ylabels at=edge right
				,xlabels at=edge bottom
               	,yticklabels at=edge right
				,xticklabels at=edge bottom
			},
    		width = 5cm, height = 2.778cm
			,xlabel near ticks
			,ylabel near ticks
			, xmin=0, xmax=0.01
			, ymin=-1.1, ymax=1.1
			, ytick={-1, 0,+1}
			, yticklabels={-1, 0, +1}
			, xtick={0.002,0.004,0.006,0.008}
		    , xticklabels={2,4,6,8}
			, grid=both
			, ylabel shift = -2 pt
			, title style={yshift=-7pt}
			, every axis xlabel/.style={}
			, ylabel style={align=center, font=\small\linespread{0.8}\selectfont}
		    , xticklabel style={
						/pgf/number format/fixed,
						/pgf/number format/precision=5
			}
			,scaled x ticks=false
			]
			\nextgroupplot[anchor=north west, at={($(left plots c3r1.east) + (0.1cm,-0.1cm)$)}, ylabel={$[0.0,0.1[$\\\textcolor{gray}{$[0.1,0.2[$}}] 
				\addplot[very thick, const plot mark left]  table [x index=0, y index=1]{data/sliding/forward_opt_x10_t10/controls/control_opt_0.dat};
				\addplot[very thick, const plot mark left,gray]  table [x index=0, y index=1]{data/sliding/forward_opt_x10_t10/controls/control_opt_1.dat};
			\nextgroupplot[ylabel={$[0.2,0.3[$\\\textcolor{gray}{$[0.3,0.4[$}}] 
				\addplot[very thick, const plot mark left]  table [x index=0, y index=1]{data/sliding/forward_opt_x10_t10/controls/control_opt_2.dat};
				\addplot[very thick, const plot mark left,gray]  table [x index=0, y index=1]{data/sliding/forward_opt_x10_t10/controls/control_opt_3.dat};
			\nextgroupplot[ylabel={$[0.4,0.5[$\\\textcolor{gray}{$[0.5,0.6[$}}] 
				\addplot[very thick, const plot mark left]  table [x index=0, y index=1]{data/sliding/forward_opt_x10_t10/controls/control_opt_4.dat};
				\addplot[very thick, const plot mark left,gray]  table [x index=0, y index=1]{data/sliding/forward_opt_x10_t10/controls/control_opt_5.dat};		
		    \nextgroupplot[ylabel={$[0.6,0.7[$\\\textcolor{gray}{$[0.7,0.8[$}}] 
				\addplot[very thick, const plot mark left]  table [x index=0, y index=1]{data/sliding/forward_opt_x10_t10/controls/control_opt_6.dat};
				\addplot[very thick, const plot mark left,gray]  table [x index=0, y index=1]{data/sliding/forward_opt_x10_t10/controls/control_opt_7.dat};
		    \nextgroupplot[ylabel={$[0.8,0.9[$\\\textcolor{gray}{$[0.9,1.0[$}}] 
				\addplot[very thick, const plot mark left]  table [x index=0, y index=1]{data/sliding/forward_opt_x10_t10/controls/control_opt_8.dat};
				\addplot[very thick, const plot mark left,gray]  table [x index=0, y index=1]{data/sliding/forward_opt_x10_t10/controls/control_opt_9.dat};
			\end{groupplot}
		\node[rotate=90,anchor=center] at ($(current bounding box.west) + (-7pt,0)$) {Height $y/\si{\milli\meter}$};
		\node[rotate=90,anchor=center] at ($(current bounding box.east) + (+7pt,0)$) {Control $Bu$};
		\node[anchor=center] at ($(current bounding box.south) + (0pt,-7pt)$) {Pos. on plate $x/\si{\milli\meter}$};
		\node[above] at ($(current bounding box.north) + (120pt,-16pt)$) {Optimal case: $R=10$, $S=10$};
		\node[above] at ($(current bounding box.north west) + (20pt,-16pt)$) {Time/s};
		\node[above,align=center, execute at begin node=\setlength{\baselineskip}{0pt}] at ($(current bounding box.north east) + (-26pt,-32pt)$) {Time\\interval/s};
		\coordinate (origo) at ($(current bounding box.south east) + (-60pt,25pt)$);
		\draw[thick] (origo) -- ++(-50pt,0) node (mary) [above]{inclination:};
		\draw[thick] (origo) -- ++(-135:30pt) node (bob) []{};
		\pic [draw,left, "$\alpha=\SI{45}{\degree}$", angle eccentricity=1.5] {angle = mary--origo--bob};
	\end{tikzpicture}
	\caption{Development of the sliding and pinned droplets over time on a plate with an inclination to the horizontal of $\alpha=\SI{45}{\degree}$: no control action with $Bu=0$ (first column), constant control action with $Bu=-0.7071$ (second column) and optimal control action for $R=10$ and $S=10$ (third column). In the fourth column the optimal control action is displayed for the time intervals. The desired shape and position of the droplet $\varphi_d$ is included as the dashed line.}	
	\label{fig:sli_results}
\end{sidewaysfigure}

The domain was of dimension $\Omega = 2l_d\times l_d$.
We assumed symmetric conditions at the left side of the domain and solid walls at the top, bottom and right side of the domain.
The inclined bottom wall contained the control of the contact angle.
The inclination of the plate was $\SI{45}{\degree}$ to the horizontal.
Exemplary, we used the dashed line in~\cref{fig:sli_results} as the desired shape $\varphi_d$ at this position on the inclined surface.
The desired droplet was produced by simulating the forward model from a cap-shaped droplet at $(x,y)=(1.25 l_d,0)$ with $Bu=-0.7071$ and $\theta_{eq}=\SI{90}{\degree}$.
In the first row of~\cref{fig:sli_results} the initial droplets at $t=0$ and $(x,y)=(0.75 l_d,0)$ with $Bu=0$ and $\theta_{eq}=\SI{90}{\degree}$ are shown.
The values for $R$ and $S$ are given in~\cref{tab:cases}.
We considered ten stripes with each ten switching points in time.
The stripes as well as the switching points were equally distributed over the space of $\SI{10}{\milli\metre}$ and the time horizon of \SI{1}{\second}.

\paragraph{No Control}
At first we show the droplet sliding down the inclined surface without any control at all ($Bu=0$ for all times), see the first column in~\cref{fig:sli_results}.
Secondly, one might naively set the control to the finally desired contact angle of \SI{135}{\degree}.
This is displayed in the second column.
In both cases the droplet is far from the desired shape as well as the desired position at the specified time.
It is clearly evident that finding the right control actions by trial-and-error might be certainly cumbersome.
Furthermore, we do not only want to obtain any controls but the controls which minimize the functional~\cref{prob:O:Ph}.

\paragraph{Optimal Control}
In the third column of~\cref{fig:sli_results} the droplet's isolines calculated from the optimal control action are displayed.
We notice, that in the optimal case we match the desired shape at the desired position very well between \SI{0.8}{\second} and \SI{1.0}{\second}.
Comparing this to the naive approaches with $Bu=0$ and $Bu=-0.7071$ the improvement is tremendous.
During the sliding and pinning the droplet exhibits multiple surprising shapes due to the strong impact of the control actions on the droplet.

The control actions $Bu$ in each time interval are shown in the last column of~\cref{fig:sli_results}
It is striking how complex these optimal controls $Bu$ are.
We would have found this very difficult and costly to find this control action by trial-and-error.
While pulled by gravity, each of the droplet's contact points are forced to recede and spread multiple times.
See for example between \SI{0.4}{\second} and \SI{0.6}{\second} the stripes between \SI{1}{\milli\metre} to \SI{2}{\milli\metre} and \SI{7}{\milli\metre} to \SI{8}{\milli\metre}. 
The control action jumps from between large and small values, which leads to the pinning of the contact points.

\section{Conclusion}
In this work we demonstrated for the first time some capabilities of optimal control of both the shape and position of a droplet.
Therefore, we considered the active control of a sliding droplet using temporally and spatially varying optimal contact angle distributions.
In this basic example common for droplet-based microfluidics the droplet slides on an inclined surface and gets pinned at a specific position with a desired shape.
In our demonstration, the final position and shape of the droplet matched the desired properties almost perfectly.
The dynamics of the droplets were calculated using a phase field model.
We solved the optimal control problem using a quasi-Newton method. 
The Gradients were calculated using the adjoints.
The resulting controls were profoundly complex and would not have been found by trial-and-error.
Our work indicates that optimal control of droplets is not only possible but might enable further automation of droplet-based microfluidic devices for example in high-throughput applications.
However, more in-depth research of the individual aspects of the concept, like objectives or constraints, is required.
This is our subject for future work.

\section*{Acknowledgments}
The first and third author thank the German Research Foundation (DFG) for the financial support within the project RE 1705/16-1.

\bibliographystyle{apalike}
\bibliography{literature}

\end{document}